\documentclass{aa}
\usepackage{graphicx}
\usepackage[varg]{txfonts}
\usepackage{natbib}
\usepackage{pst-eps}
\usepackage{combelow}

\newcommand{\zav}[1]{\left(#1\right)}

\newlength\staretab

\newcommand\de{\text{d}}

\newcommand\kms{\ensuremath{\text{km}\,\text{s}^{-1}}}

\newcommand{\hvezda}{EPIC~206197016}

\begin{document}

\title{EPIC 206197016: A very hot white dwarf orbited by a strongly irradiated
red dwarf\thanks{Based on
observations collected at the European Southern Observatory, Paranal, Chile (ESO
programme 0103.D-0194(A)).}}

\author{J.~Krti\v{c}ka\inst{1} \and A.~Kawka\inst{2} \and 
        Z.~Mikul\'a\v sek\inst{1} \and L.~Fossati\inst{3} \and
        I.~Krti\v ckov\'a\inst{1} \and M.~Prv\'ak\inst{1} \and
        J.~Jan\'\i k\inst{1} \and R.~Liptaj\inst{1} \and
        M.~Zejda\inst{1} \and E.~Paunzen\inst{1}}

\institute{Department of Theoretical Physics and Astrophysics, Faculty of
           Science, Masaryk University, Kotl\'a\v rsk\' a 2, Brno,
           Czech Republic \and
           International Centre for Radio Astronomy Research,
           Curtin University, Perth, Australia \and
           Space Research Institute, Austrian Academy of Sciences,
           Schmiedlstrasse 6, Graz, Austria 
}

\date{Received}

\abstract
{Very precise satellite photometry has revealed a large number of variable
stars whose variability is caused either by surface spots or by binarity.
Detailed studies of such variables provide insights into the physics of these
objects.}
{We study the nature of the periodic light variability of the white
dwarf EPIC 206197016 that was observed by the K2 mission.}
{We obtain phase-resolved medium-resolution spectroscopy of EPIC 206197016 using XSHOOTER
spectrograph at VLT to understand the nature of the white dwarf variability. We
use NLTE model atmospheres to determine stellar parameters at individual
phases.}
{EPIC 206197016 is a hot DA white dwarf with $T_\text{eff}=78\,$kK. The
analysis of the spectra reveals periodic radial velocity variations that can
result from gravitational interaction with an invisible secondary whose mass corresponds to a red dwarf. The close proximity of the two stars where the
semimajor axis is about $3\,R_\odot$ results in the irradiation of the companion
with temperatures more than twice as high on the illuminated side compared
to the nonilluminated hemisphere. This effect can explain the observed light
variations. The spectra of the white dwarf show a particular feature of the
Balmer lines called the Balmer line problem, where the observed cores of the
lower Balmer lines are deeper than predicted. This can be attributed to
either weak pollution of hydrogen in the white dwarf atmosphere by heavy elements or to the
presence of a circumstellar cloud or disk.}
{}



\maketitle

\section{Introduction}

In a multiple stellar system, extrinsic light variability typically appears due
to geometrical reasons, either as a result of stellar rotation or orbital
motion. The periodic nature of this variability enables us to learn about the
geometry of a studied object that would otherwise probably remain unresolved.

Rotational light variability is very common among stars of different spectral
types. From cool to hot, many stars show photometric spots or patches, which
cause light variability as a result of stellar rotation
\citep[e.g.,][]{chodkyne,hummer,kunites,mobstersi,moma}. However, the physical
nature of photometric spots may differ for stars with different spectral types.
While the presence of spots in cool stars is connected with the convective flux
being suppressed in regions with strong magnetic fields \citep{iss,jadav},
patches on hot stars appear most frequently as a result of flux redistribution
in abundance spots \citep{peter,vlci,ministr}.

Light variations due to rotation have also been detected in white dwarfs.
\citet{dupchaven} found periodic extreme ultraviolet variability in the DA white
dwarf GD~394, which was attributed to a silicon surface spot. More recently,
\citet{wiljakoven} detected optical variations of this star from Transiting
Exoplanet Survey Satellite (TESS) observations. In many cases, the nature of
light variability of white dwarfs is not reliably known
\citep{kurtis,maoz,kockov}. Nevertheless, apart from chemical spots
\citep{kilic}, magnetic fields are often considered a main cause
\citep{cerving}.

Hot white dwarfs may have surface spots as a result of radiative diffusion,
which affects their outer layers \citep{unbu}. White dwarfs accrete debris
coming from their planetary systems \citep[e.g.,][]{johnbt}, which may also lead
to uneven distribution of metals across their surfaces. In strongly magnetized
white dwarfs, even continuum absorption coefficients may be affected by the
magnetic field \citep{kemp,canthom,javor}, providing another possible source
of light variability.

However, there are other mechanisms that may mimic the variability due to the
surface spots and hamper the search for surface inhomogeneities. In many cases,
variations due to binary effects, that is, ellipsoidal variations and reflection
effects, may resemble spot variability \citep[e.g.,][]{green15k}. The
ellipsoidal variability appears due to tidal deformation of surfaces of binary
components \citep{kopal}. This is therefore particularly important to consider
when the binary separation is comparable to the radii of the individual
components, for instance, in main-sequence binaries with periods of few days
\citep[e.g.,][]{koueli,koceli} or in subdwarf binaries with periods on the order
of tens of minutes \citep[e.g.,][]{adsub,kupeli}. The so-called reflection
effect appears to be due to mutual irradiation of components \citep{kopal}, and
it becomes especially important in binaries with large effective temperature
differences, such as in systems containing white dwarfs. For instance,
reflection or reprocessing effects are seen in many close binaries where the
companion of a white dwarf is either a low-mass main-sequence star
\citep{adelareflex} or a brown dwarf \citep{dobrakase}. For short-period
systems, the light curves are modified by relativistic effects \citep[so-called
Doppler beaming,][]{logaudi}.

To resolve these issues and to understand the nature of light variability of
white dwarfs, we inspected the list of white dwarfs observed by the Kepler
satellite \citep{hermes}, looking for those that may show periodic variability.
We selected the most promising ones for a detailed spectroscopic study to
identify their nature \citep{esobtprmhvi}. In this paper, we report the detailed analysis
of \hvezda\ (WD 2244-100, PB 7199, SDSSJ224653.72-094834.4).

\section{Observations}

The available astrometric and photometric data for \hvezda\ are summarized in
Table~\ref{hveztab}. The coordinates and $R$ magnitude were obtained from the
Mikulski Archive for Space Telescopes (MAST) K2 catalog \citep{epic}. The
distance $d$ was determined using the parallax from the Gaia Data Release 3
(DR3) data \citep{gaia1,gaia2,gaia3}.

We performed phase-resolved spectroscopy of \hvezda\ as part of the ESO proposal
0103.D-0194(A). The spectra were acquired with the XSHOOTER spectrograph
\citep{xshooter} mounted on the 8.2m UT2 Kueyen Telescope. The spectroscopic
observations are summarized in Table~\ref{spek}. We utilized the UVB and VIS
arms, which provide an average spectral resolution ($R=\lambda/\Delta\lambda$)
of 9700 and 18\,400, respectively.

The spectra were manually calibrated and shifted to the rest frame using the
radial velocities $v_\text{rad}$ (given in Table~\ref{spek}) determined from
each spectrum by means of the cross correlation technique. We used a
theoretical spectrum as a template \citep{zvezimi}. In addition, we  tested the
effect of the emission peak inside the H$\alpha$ profile on the determined radial
velocities, but we found it to be negligible.

\begin{table}[t]
\caption{Astrometric and photometric properties of \hvezda\ (top) and
derived mean parameters (bottom).}
\label{hveztab}
\centering
\begin{tabular}{lc}
\hline
$\alpha$ (J2000)   & 22h 46m 53.727s \\
$\delta$ (J2000)   & $-9^\circ$ $48'$ $34.51''$ \\
$\pi$ [mas]        & $ 1.902\pm0.079$ \\
$d$ [pc]           & $526\pm22$ \\
$m_R$ [mag]        & $16.671\pm0.004$  \\
\hline
$T_\text{eff}$ [K]      & $78\,000\pm3000$ \\
$\log g$ [cgs]          & $7.51\pm0.10$ \\
$M$ $[M_\odot]$         & $0.57$ \\
$R$ $[R_\odot]$         & $0.021\pm0.001$ \\
$\log\varepsilon_\text{He}$ & $<-2$ \\
\hline
\end{tabular}
\end{table}

The K2 photometry of \hvezda\ was obtained from the MAST
archive\footnote{http://archive.stsci.edu} \citep{epic}. The photometry shows
periodic light variations, from which we determined the ephemeris (in barycenter
corrected Julian date)
\begin{equation}
\label{hvezdaper}
\text{BJD}=2\,457\,011.2033(14)+0.829101(23)E
\end{equation}
using the method described by \citet{mikmon}. Here, the zero phase corresponds to
the maximum of the light curve.

\begin{table*}[t]
\centering
\caption{List of \hvezda\ spectra used for the analysis.}
\label{spek}
\begin{tabular}{lcccccc}
\hline\hline
Arm & Spectrum & BJD$-2\,400\,000$\tablefootmark{$\ast$} & Phase\tablefootmark{$\ast\ast$}&
Exposure time [s] & S/N\tablefootmark{$\ast\!\ast\!\ast$} &
$v_\text{rad}$\tablefootmark{$\dagger$} [km$\,$s$^{-1}$] \\
\hline
UVB & XSHOO.2019-07-07T09:19:13.007 & 58671.90734 & 0.018 & 2700 & 50 & $2\pm10$ \\ 
VIS & XSHOO.2019-07-07T09:20:00.001 & 58671.90734 & 0.018 & 2606 & 20 & \\ 
UVB & XSHOO.2019-07-14T04:29:41.006 & 58678.70679 & 0.219 & 2700 & 80 & $19\pm30$ \\ 
VIS & XSHOO.2019-07-14T04:30:28.010 & 58678.70679 & 0.219 & 2606 & 40 & \\ 
UVB & XSHOO.2019-09-01T04:42:29.041 & 58727.71759 & 0.332 & 2700 & 70 & $28\pm12$ \\ 
VIS & XSHOO.2019-09-01T04:43:16.002 & 58727.71759 & 0.332 & 2606 & 30 & \\ 
UVB & XSHOO.2019-08-09T02:11:05.004 & 58704.61200 & 0.464 & 2700 & 50 & $15\pm10$ \\ 
VIS & XSHOO.2019-08-09T02:11:52.008 & 58704.61200 & 0.464 & 2606 & 20 & \\ 
UVB & XSHOO.2019-09-03T01:55:56.005 & 58729.60193 & 0.605 & 2700 & 70 & $27\pm10$ \\ 
VIS & XSHOO.2019-09-03T01:56:43.009 & 58729.60193 & 0.605 & 2606 & 30 & \\ 
UVB & XSHOO.2019-08-30T03:23:06.009 & 58725.66246 & 0.853 & 2700 & 70 & $-7\pm7$ \\
VIS & XSHOO.2019-08-30T03:23:53.003 & 58725.66246 & 0.853 & 2606 & 30 & \\ 
\hline
\end{tabular}
\tablefoot{\tablefoottext{$\ast$}{Mid-exposure time.}
\tablefoottext{$\ast\ast$}{Determined from photometry.}
\tablefoottext{$\ast\!\ast\!\ast$}{At 4400\,\AA\ and 6500\,\AA\ for the UVB and VIS arms,
respectively.}\tablefoottext{$\dagger$}{Average from both arms.}}
\end{table*}

The observational data were supplemented with photometry derived using the
Virtual Observatory SED
Analyzer\footnote{http://svo2.cab.inta-csic.es/theory/vosa/} web tool
\citep[VOSA;][]{vosa}, which was used to build the spectral energy distribution. We used
photometry from Galaxy Evolution Explorer \citep[GALEX;][]{vosagalex}, Sloan
Digital Sky Survey \citep[SDSS;][]{vosasloan}, the Carlsberg Meridian Telescope
CCD drift scan survey \citep{vosaapass}, the Panoramic Survey Telescope and
Rapid Response System 1 survey \citep[Pan-STARRS1;][]{vosapan} , Gaia
\citep{gaia2}, Visible and Infrared Survey Telescope for Astronomy survey
\citep[VISTA;][]{vosavista}, and Wide-field Infrared Survey Explorer
\citep[WISE;][]{vosawise}.

\section{Spectroscopic analysis}

\subsection{Stellar parameters}

\begin{figure*}[t]
\centering
\includegraphics[width=\hsize]{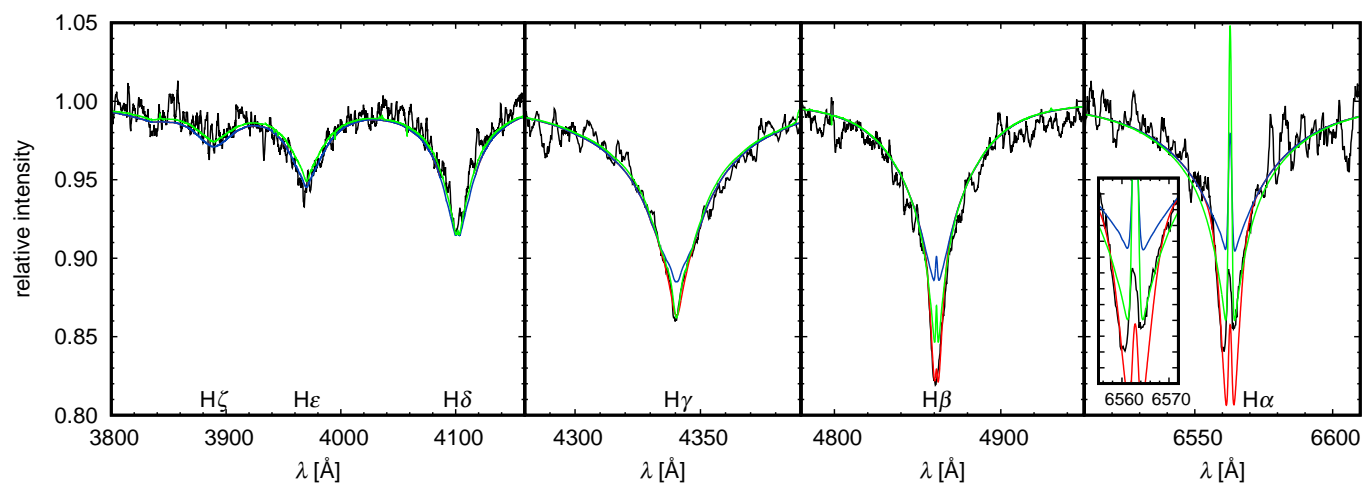}
\caption{Comparison of observed and fitted spectra. The black curve denotes
observed spectrum ($\varphi=0.332$), the blue curve corresponds to the fitted
spectrum, the red curve denotes the spectrum with additional magnetospheric
absorption, and the green curve denotes the spectrum from the model that
includes heavy elements. The inlet shows the central part of the H$\alpha$
profile.}
\label{K206197016_1_UVB}
\end{figure*}

We fitted the stellar spectra to derive the stellar parameters. We used the
TLUSTY NLTE model atmospheres and the SYNSPEC code \citep{ostar2003,bstar2006}
to calculate a grid of synthetic spectra. We interpolated  within calculated
synthetic spectra to get the best match between theoretical and observed
spectra. Visual inspection did not reveal the presence of any helium lines and
pointed to a very high temperature of the star. Therefore, we selected a grid
parameterized by the effective temperature $T_\text{eff}\in[70,\,80,\,90]\,$kK,
surface gravity $\log g\in[7.00,\,7.25,\,7.50,\,7.75]$, and helium abundance
$\varepsilon_\text{He}=\log(N_\text{He}/N_\text{H})\in[10^{-3},\,10^{-2}]$ to
obtain a grid of 24 NLTE models of stellar atmospheres and synthetic spectra. We
did not include any elements heavier than helium into our calculations. We used
both normalized and flux calibrated spectra, which gave nearly the same results.

An example of the fitted spectra is given in Fig.~\ref{K206197016_1_UVB}, which
shows that the models are able to nicely reproduce the higher order Balmer lines
and the wings of the lower order Balmer lines, while the observed cores of
H$\alpha$, H$\beta$, and H$\gamma$ are deeper than predicted. This is a
manifestation of the so-called "Balmer line problem"
\citep{napibal,werbal,werakbal} wherein the observed lower Balmer lines
appear stronger than the higher Balmer lines. Therefore, the lower Balmer lines
can be fitted with a lower effective temperature than the higher ones. The
origin of this problem is not fully clear, but it can possibly
be attributed to metal line blanketing or to the presence of a circumstellar
magnetosphere \citep{werbal,venbal,wrr}. To mitigate the problem, we fit only
the spectra from the UVB arm.

The final determined parameters are given in Table~\ref{hveztab}. They were
determined as the average of the parameters derived from the individual UVB
spectra. The derived parameters support our initial estimate that the object is
a very hot DA white dwarf with no traces of helium in the atmosphere. The
estimated temperature is somewhat lower than the value of
$100\,000\pm4000\,\text{K}$ determined by \citet{tresloan} using similar
methodology as we used here. The difference in temperature is perhaps connected with the fact
that their analysis is based on spectra of lower quality.

\begin{figure}
\includegraphics[width=0.5\textwidth]{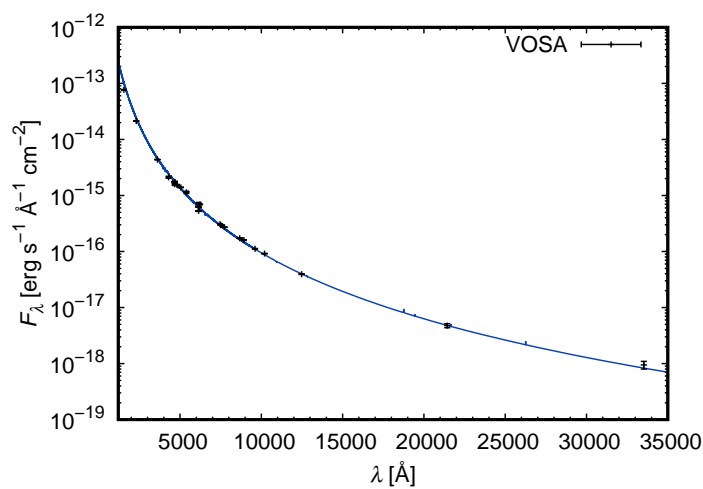}
\caption{Comparison of the predicted spectral energy distribution calculated for
the determined parameters (Table~\ref{hveztab}, solid blue line) with
observational data derived using the VOSA web tool \citep{vosa}.}
\label{206197016sed}
\end{figure}

Since the line cores are most significantly affected by the Balmer line problem,
we separately fit all observed Balmer lines (including H$\alpha$) while only
considering the line wings. This gives slightly more consistent parameters of
$T_\text{eff}=83\pm3\,$kK and $\log g=7.64\pm0.07$. However, these values are
(within the errors) the same as those derived from the fitting of whole line
profiles. This likely implies that the Balmer line problem does not
significantly affect the derived stellar parameters. Furthermore, the good
agreement between predicted (neglecting interstellar reddening) and observed
spectral energy distribution shown in Fig.~\ref{206197016sed} also supports the
reliability of the derived parameters. However, the spectral energy distribution
provides just a loose constraint on the effective temperature because the fit
performed using a larger grid of model atmosphere fluxes from \citet{levyhag}
yields only $T_\text{eff}>40\,\text{kK}$. The fit of the spectral energy
distribution with temperature from spectroscopy gives a radius estimate of
$0.021\pm0.001\,R_\odot$. With $\log g$ derived from spectroscopy
(Table~\ref{hveztab}), this gives a stellar mass of $0.52\pm0.12\,M_\odot$,
which is consistent with the value derived in the following analysis from
evolutionary tracks within the errors. The lack of any infrared excess implies
that any potential companion should be of a spectral type later than M8, as
derived using the stellar parameters from \citet[or extrapolating data from
\citealt{novyhar}]{har}. Because the spectral energy distribution is rather
insensitive to the effective temperature of the white dwarf at observed
wavelengths, this conclusion is not affected by observational uncertainties of
derived parameters.

Besides the Balmer lines, we have not found any clear evidence for the presence
of other stellar lines in the spectra. We found absorption lines at 3384\,\AA\
(identified as \ion{Ti}{ii}), 3934\,\AA, and 3968\,\AA\ (\ion{Ca}{ii}), which
are narrow and therefore likely  have interstellar (or circumbinary) origin.
Their origin is further supported by our analysis showing that these lines are
stationary and do not change their wavelength. Furthermore, the spectra display
a broad absorption feature at about 4606\,\AA, which we were unable to clearly
identify.

From the derived effective temperature and surface gravity, we determined a mass
of $0.57\,M_\odot$ and a cooling age of $2.37 \times 10^5$\,yr using DA cooling
tracks from the Montreal
group\footnote{http://www.astro.umontreal.ca/\~{}bergeron/CoolingModels}
\citep{holbe,beda}. The derived mass nicely agrees with
$M_1=0.59\pm0.02\,M_\odot$, determined by \citet{tresloan}. We also checked the
distance using the obtained values and the SDSS $g$ band photometry and found it
to be $560\pm65$\,pc, which is consistent with the Gaia distance
(Table~\ref{hveztab}), within the uncertainties.

\subsection{Radial velocity curve and evolutionary status}

\begin{figure}[t]
\centering
\includegraphics[width=\hsize]{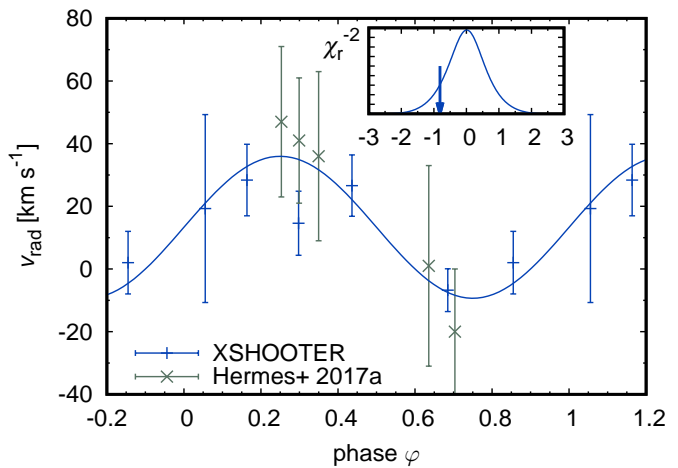}
\caption{Phase-folded XSHOOTER (blue) and \citet[green]{divokyhermes} radial
velocity values. The data were phased with period from radial velocity
variations. The inlet shows the periodogram of the radial velocity values as a
function of the relative period difference $\Delta P/P\times10^4$, where $P$ is
the orbital period derived from the radial velocity variations. The arrow
indicates the period derived from the photometry. }
\label{206197016vr}
\end{figure}

\citet{divokyhermes} detected radial velocity variations and suspected that
\hvezda\ is a binary. We combined their radial velocity measurements with
our determinations (Table~\ref{spek}) in Fig.~\ref{206197016vr}. The best
phasing of all radial velocities can be achieved with a period of
$P=0.829168(54)\,$d, which is slightly higher than that derived from photometry.
The assumption of circular orbits provides a good fit to the radial velocity
curve with semiamplitude $v_1=23\pm4\,\kms$. The mass ratio is given by
\begin{equation}
\label{keplerdopler}
\zav{1+\frac{M_1}{M_2}}^3\zav{1+\frac{M_2}{M_1}}^{-1}=\frac{2\pi GM_1}{P v_1^3}
\sin^3 i,
\end{equation}
where $i$ is the orbital inclination. With a white dwarf mass of
$M_1=0.57\,M_\odot$ and the derived semiamplitude of the radial velocities, the
solution of Eq.~\eqref{keplerdopler} gives a secondary mass of
$M_2\geq0.06\,M_\odot$. This corresponds to either a low-mass red dwarf or a
high-mass brown dwarf. The orbital separation estimated from the third Kepler law is
$a>3.1\,R_\odot$, which is large enough to accommodate any cool main-sequence
star into the system.

Comparable white dwarf binaries with a brown dwarf or a low-mass red dwarf
companion have already been found \citep{bezucke,kavenbin,fari,burst,kosak}. The
initial-final mass relation of \citet{cummif} estimates an initial mass of about
$1.0\,M_\odot$ for a white dwarf mass of $M_1=0.57\,M_\odot$. However, the real
initial mass may be significantly higher because this estimate does not take
into account previous binary interaction. The stellar radius during late
evolutionary phases of solar-mass stars exceeds the current separation of binary
components by orders of magnitude, therefore indicating that the system has
undergone a common envelope phase \citep{paspol,zormat}.

We explored the possible evolutionary history of the system using the output
from the Binary Population and Spectral Synthesis code (BPASS; \citealt{stel}
version 2.2.1). We searched the available tracks for the system with parameters
that are close to the observed values. We selected a model with the initial
parameters $M_1=3\,M_\odot$, $M_2=0.3\,M_\odot$, and $P=6.3\,$d. This system
passes through a contact binary phase, during which the primary loses a
significant fraction of its mass \citep[e.g.,][]{kram} and becomes a helium
dominated subdwarf with a mass of $0.42\,M_\odot$ that likely evolves into the
realm of white dwarfs (this is not covered by the available track). There is
still some hydrogen left on the stellar surface, and consequently, one can
assume that helium later sinks down and the object appears as a hydrogen
dominated DA white dwarf \citep{bbb}. The orbital period of the system shrinks
to about $0.6\,$d. The grid of evolutionary tracks does not contain systems with
mass ratios lower than $0.1$. Consequently, we were unable to find a model with
a secondary mass close to the value estimated from the observations. However, a
comparison with a track that has a higher mass of the secondary star showed that
the final parameters of the primary are rather insensitive to the initial mass
of the secondary. Therefore, although the final parameters do not perfectly
match those of \hvezda, it is reasonable to assume that a system with a lower
initial mass of the secondary would evolve in a similar way and explain the
final parameters better.

The binary might have experienced even more complex evolution featuring
a second common envelope stage after the secondary left the main sequence
\citep{stabzen}. However, this seems to be less likely, as the light variations
studied in the next section do not indicate that the secondary left the main
sequence.

\section{Nature of the light variations}

\begin{figure}[t]
\centering
\includegraphics[width=\hsize]{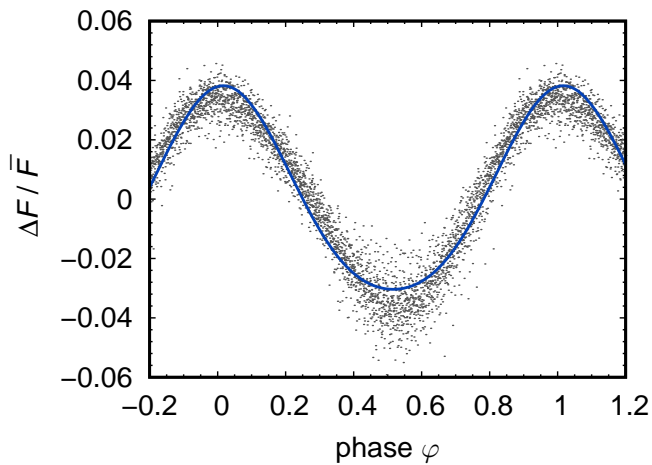}
\caption{Phase-folded light curve derived from K2 photometry. We plotted the
relative difference of the observed flux from the mean value. The solid blue
line is the modeled variability computed accounting for irradiation.}
\label{206197016prom}
\end{figure}

Although our test using the spot model of \citet{prvakphd} revealed that the
observed light variations can be caused by spots, given the alleged binary
nature of the object, it is more reasonable to assume a binary origin of the
light variations for \hvezda. In this case, the light variations could either
be due to the distortion of the white dwarf surface or to reflection.

If the light variations are due to distortion of the white dwarf surface
(ellipsoidal variability), then the orbital period would be twice that
determined in Eq.~\eqref{hvezdaper}. By using Kepler's third law, the semimajor
axis would be at least $4.9\,R_\odot$. To estimate the amplitude of the
ellipsoidal variability, we used our code \citep{esobtprmhvi}, which within the
Roche model predicts the light curve of the primary due to distortion of its
surface by an accompanying secondary. Assuming a companion mass of
$0.2\,M_\odot$, the derived amplitude is three orders of magnitude lower than
the observed amplitude of the light variability. Consequently, we conclude that
the ellipsoidal variations due to distortion of the white dwarf are unlikely to cause
light variability in \hvezda.

Another possibility is that the light variations are due to a reflection effect
on a white dwarf companion \citep[e.g.,][]{budref,scherebos}. In such a case, the
orbital period would be equal to that determined in Eq.~\eqref{hvezdaper}. For
the modeling of the \hvezda\ light curve, we used our own code that calculates the
light curve of the binary with irradiation. The code assumes spherical shape of
both stars, which is reasonable due to the small radii of both components in
comparison to the binary separation. We assumed that both components radiate
as black bodies and that the limb darkening can be described by Lambert's cosine
law \citep{mikrobalian}. The code integrates the flux over the whole
secondary surface to get the magnitude of binary in a given phase. We adopted an
albedo of $A=\frac{1}{2}$, which is reasonable for cool stars with deep
convective envelopes \citep{rucalb}. We assumed a general elliptical orbit. The
resulting light curve typically resembles a sinusoidal with sharper maxima and
shallower minima \citep{barfour}. Computed light curves favorably compare
with data derived in the literature for other systems
\citep{testodraz2,testodraz1}.

We fit the observed light curve as a function of binary and secondary
parameters, that is, the orbital inclination, eccentricity $e$, effective
temperature $T_2$, radius $R_2$, and mass $M_2$ of the secondary. We fixed both
the white dwarf parameters (given in Table~\ref{hveztab}) and orbital period.
However, the solution was not well constrained given the large number of free
parameters. Therefore, we additionally fixed the secondary effective temperature
to a value roughly corresponding to the maximum one allowed by the spectral
energy distribution. Using the evolutionary models of low-mass stars by
\citet{barnov}, this yielded an estimate of the mass of the secondary. Along
with Eq.~\eqref{keplerdopler}, this constraint allowed us to fix the orbital
inclination. The resulting fit of the light curve (Fig.~\ref{206197016prom})
gave secondary and orbital parameters $T_2=3000\,\text{K}$, $M_2=0.12\,M_\odot$,
$R_2=0.08\,R_\odot$, $i=40^\circ$, and $e=0$. The derived parameters of the
secondary are reasonably close to the values derived from the evolutionary
tracks of \citet{barnov}, which predict an evolved $0.11\,M_\odot$ mass star
with a radius of $0.135\,R_\odot$ and effective temperature of $2900\,\text{K}$.
Based on these results, it follows that the secondary is most likely a red
dwarf.

As a result of irradiation, the effective temperature of the cool companion
increases from about 3\,kK to nearly 6\,kK on its illuminated side. This
significantly changes the atmospheric structure \citep{pet,lotca,sluneni}, likely
dissociating molecules and enhancing the atmospheric scale height \citep{hubif}.
A large temperature difference between both hemispheres causes departures from
hydrostatic equilibrium, resulting in a flow similar to the geostrophic flow on
Earth. These effects are further modified by the  energy stored in hydrogen
dissociation and recombination \citep{tankoma}. This might explain the slight shift between predicted and observed light minima
(Fig.~\ref{206197016prom}).

We did not find any signatures of the secondary in the observed spectral
energy distribution (Fig.~\ref{206197016sed}) nor in the spectra. The secondary can
be expected to mainly dominate in the VIS part of the spectrum, which has a low signal-to-noise ratio (Table~\ref{spek}). This perhaps explains the missing spectral
signatures of the secondary. The emission lines of the secondary could appear in the
H$\alpha$ line (e.g., \citealt{adeladvoj,adelareflex}), which is indeed
variable. However, the emission also has contribution from the white dwarf
itself (Fig.~\ref{K206197016_1_UVB}), which makes the analysis of this line
problematic.

\section{The origin of the Balmer line problem}

The cores of the observed profiles of the H$\alpha$ and H$\beta$ lines of \hvezda\ are
stronger than theory predicts. This outcome is called the Balmer line problem
\citep{napibal,werbal,werakbal}. To test possible periodic modulation of the
Balmer line problem, we calculated the difference between the equivalent widths
of predicted and observed spectra. The Balmer line core may indeed show some
variability, but we consider this to be inconclusive due to a lack of correlation
between variations in individual lines.

\subsection{Second component}

We tested if the Balmer line problem could not be explained by a combination of
hot white dwarf spectrum with spectrum of another star. We tested for the
presence of an additional white dwarf with stellar parameters similar to those
derived for the primary star; a hot OB star, using spectra from OSTAR2002 and
BSTAR2006 grids \citep{ostar2003,bstar2006}; and a cooler star with an effective
temperature in the range of $3500-15\,000\,$K, using the ATLAS9 grid
\citep{atlas9}. The inclusion of an additional component did not lead to a
significant improvement of the fit. The reason for this is that the optical
region corresponds to the Rayleigh-Jeans tail of the spectra, and the
combination of two spectra with similar shapes does not provide better agreement
with observations. The further inclusion of a second light led  to the
appearance of numerous lines of the secondary \citep{veneuve,dvasyn}, which do
not appear in the observed spectra.

\subsection{Heavy elements}

Line-driven winds fade out during evolution along the white dwarf cooling track
\citep{btvit}. Their disappearance opens a door for gravitational settling,
which removes heavy elements from white dwarf photospheres \citep{unbu,un} in
cases where radiative levitation \citep{chadif} is inefficient. However, traces
of heavy elements may still remain in the atmospheres of hot white dwarfs,
possibly influencing the temperature distribution and emergent spectra. This
could explain the difference between predicted and observed Balmer lines
\citep{werbal,venbal,wrr}.

To test this possibility, we prepared an additional grid of helium-free model
atmospheres and synthetic spectra with $T_\text{eff}\in[80,90]\,$kK and $\log
g\in[7.50,7.75]$. We adopted a reduced mass-fraction of heavy elements
$Z=0.1Z_\odot$ and fitted individual spectra with these new models. Because the
presence of metals affects mostly the line cores, the derived atmospheric
parameters agree within uncertainties with parameters derived from pure H-He
models. The resulting spectrum (Fig.~\ref{K206197016_1_UVB}) shows much better
agreement with the observations, especially in regard to the H$\alpha$ and
H$\gamma$ lines. There remains only a slight disagreement in the H$\beta$ line.

The presence of metals in the atmospheres of hot white dwarfs can be revealed
from ultraviolet observations. However, the inclusion of metals does not
fully remove the Balmer line problem and points to yet another effect
influencing the Balmer line profiles \citep{wrr}.

\subsection{Magnetospheric origin}

\citet{wrr} attributed the Balmer line problem in the hot DA white dwarf
PG~0948+534 to light absorption in the magnetosphere. Indeed, a significant
fraction of white dwarfs are found to be magnetic \citep[see][for a review]{ab}.
As a result of centrifugal force, the outflowing matter may become trapped in
the corotating magnetosphere in the regions of potential minima along the
individual field lines \citep{prus,towo}. The trapped magnetospheric matter may
cause photometric and spectroscopic variability \citep{labor,towog}. The matter
is distributed within the corotating disk, which is warped depending on the
magnetic field inclination. The inner disk radius is roughly given by the
Keplerian radius \citep{prus}, 
\begin{equation}
\label{prazak}
R_\text{K}=\zav{\frac{GMP^2}{4\pi^2}}^{1/3}
\approx 0.5\,R_\odot\zav{\frac{M}{1\,M_\odot}}^{1/3}
\zav{\frac{P}{1\,\text{h}}}^{2/3}.
\end{equation}
From this follows that white dwarfs with typical radii $R\approx0.02\,R_\odot$
should rotate with periods significantly shorter than one hour in order to have their
Keplerian radii close to the stellar radii.

The circumstellar magnetosphere extends up to the Alfv\'en radius, where the
magnetic field energy density ceases to dominate over the gas kinetic energy
density. For a stellar wind outflow, the Alfv\'en radius is given by
\citep{udorot}
\begin{equation}
\label{alfi}
{R_\text{A}}/{R_\ast}\approx0.3+(\eta_\ast+0.25)^{1/4},
\end{equation}
where the wind magnetic confinement parameter \citep{udo} is
\begin{equation}
\label{eta}
\eta_\ast=\frac{B_\text{eq}^2R_\ast^2}{\dot Mv_\infty}.
\end{equation}
Here, we assumed that the magnetic field has a dipolar topology with equatorial
magnetic field strength $B_\text{eq}$. The symbols $\dot M$ and $v_\infty$ are the wind
mass-loss rate and terminal velocity, respectively.
For a large confinement $\eta_\ast\gg1$, the Alfv\'en radius is 
\begin{multline}
\label{streda}
R_\text{A}=\frac{B_\text{eq}^{1/2}R_\ast^{3/2}}{\zav{\dot Mv_\infty}^{1/4}}
\approx9\times10^{-3}\,R_\odot\\*\zav{\frac{B_\text{eq}}{1\,\text{kG}}}^{1/2}
\zav{\frac{R_\ast}{0.01\,R_\odot}}^{3/2}
\zav{\frac{\dot M}{10^{-10}\,M_\odot\,\text{yr}^{-1}}}^{-1/4}
\zav{\frac{v_\infty}{10^3\,\kms}}^{-1/4}.
\end{multline}
This shows that in addition to a fast rotation, a magnetic field with strength
on the order of at least 100\,kG is needed to confine the material in
a sufficiently large magnetosphere.

The region around the potential minima along a given field line is filled with
magnetospheric matter with a density given by a Gaussian distribution \citep{towo}
\begin{equation}
\label{hustdisk}
\rho(\Delta s)=\rho_\text{m}\exp\zav{-\frac{\Delta s^2}{h^2}},
\end{equation}
where $h$ is the characteristic scale height and $\Delta s$ is the distance from
the potential minima along the field lines. The optical depth along a radial ray
parallel with the disk is
\begin{equation}
\label{taudi}
\tau=\int\kappa\rho\,\de l=fD\exp\zav{-\frac{\Delta s^2}{h^2}}\phi(\lambda),
\end{equation}
where $f$ is the line oscillator strength, the parameter $D$ is constant for
the lines coming from the same series, and $\phi(\lambda)$ is the line profile
(assumed to be Gaussian). The height of the disk that effectively blocks the
stellar radiation is given by the condition of unity optical depth
$\tau=1$, that is,
\begin{equation}
\Delta s\sim h\sqrt{\ln(f)+\ln(D)}.
\end{equation}
This means that the fraction of the stellar surface obscured by the disk depends
on the oscillator strength of a given line. For Balmer lines, the obscured
fraction is the strongest for H$\alpha$, and it decreases with an increasing main
quantum number of the upper level.

The absorption may vary with rotational phase in the case when the magnetic
field axis is tilted with rotational axis. We did not detect any clear
signature of photometric or spectroscopic variability in \hvezda\ that could be
attributed to the corotating magnetosphere. This could mean that the magnetic
field axis nearly coincides with the rotational axis and that we can assume the same
amount of absorption in all spectra.

We fit \hvezda\ spectra assuming additional absorption with optical depth given
by Eq.~\eqref{taudi}. The resulting spectra are overplotted in
Fig.~\ref{K206197016_1_UVB}. Apparently, the inclusion of additional absorption
significantly improves the spectral fit in the H$\beta$ and H$\gamma$ lines. The
core of the H$\alpha$ line is still not yet reproduced, what may be connected to
the presence of additional emission appearing in this line due to the disk
\citep{towog}. The magnetosphere rotates rigidly, and therefore the rotational
velocity of the material trapped in the magnetosphere increases proportionally
to the radius. Instead, the line of sight velocity of the material that obscures
the stellar limb is inversely proportional to the radius. As a result, the width
of the absorption line profile directly gives a line of sight projection of the
rotational velocity of the white dwarf. With $d=4\,$\AA, this implies $v\sin
i=300\,\kms$. With a white dwarf radius of $0.021\,R_\odot$, this gives a
rotational period of about five minutes and a Keplerian radius of about
0.07\,$R_\odot$, based on Eq.~\eqref{prazak}. Accounting for spatial scaling,
this Keplerian radius is comparable to the extension of magnetospheres of
chemically peculiar stars \citep{malykor,shalfa}.

\hvezda\ is too cool to have any wind containing hydrogen \citep{btvit}.
However, because the cooling time of hot white dwarfs \citep{milbe} is
comparable to the characteristic time of emptying the magnetosphere
\citep{towo}, the magnetospheric matter may be the remainder of line-driven wind
from previous evolutionary phases.

The magnetic field strength required to confine the matter in the magnetosphere
of white dwarfs is on the order of 100\,kG (Eq.~\eqref{streda}). Fields with
comparable strength typically leave signatures in the spectra of white dwarfs
\citep[e.g.,][]{pobliz}. However, there may be another source of the
circumstellar disk matter. The disk could be the remainder of previous interaction
phases, or it may have been formed by the matter of a former exoplanetary system
\citep{jura}. Therefore, the deep cores of the Balmer lines may originate in the
accreting Keplerian disk around the white dwarf. As Keplerian disks also
have a Gaussian density distribution in the direction perpendicular to the disk
plane, the same model as outlined for the magnetospheric matter may describe the
light absorption from the circumstellar disk.


\section{Conclusions}

We carried out a detailed spectroscopic and photometric investigation of DA
white dwarf \hvezda\ to understand the nature of its periodic light variability.
From pure hydrogen optical spectrum we determined a relatively high effective
temperature of $78\pm3$\,kK, implying that the white dwarf ascended the white dwarf cooling branch relatively recently.

The optical spectrum shows a periodic Doppler shift, which can be interpreted as being the
result of a binary motion caused by an unseen companion, presumably a red dwarf or
a brown dwarf. The reflection effect on a red dwarf would also explain the
light variability of this system.

The cores of lower Balmer lines show an unusually strong absorption, which we
were unable to reproduce with purely hydrogen model atmospheres. We failed to
pinpoint an exact cause of this effect, but the metal pollution of the
atmosphere or a circumstellar environment, either in the form of an accretion
disk or a corotating magnetosphere, could account for the enhanced absorption
reasonably well. At such high effective temperatures, detecting metals and
measuring their abundances would require ultraviolet spectra, but this would not
affect the estimate of the atmospheric parameters.

\begin{acknowledgements}
We thank Dr.~J.~Budaj and the anonymous referee for the comments that helped us 
to improve the paper. Computational resources were provided by the e-INFRA CZ
project (ID:90140), supported by the Ministry of Education, Youth and Sports of
the Czech Republic.
\end{acknowledgements}

\bibliographystyle{aa}
\bibliography{papers}

\end{document}